# The Burr XII Distribution Family and the Maximum Entropy Principle: Power-Law Phenomena Are Not Necessarily "Nonextensive"

**F. Brouers**

Institute of Physics, University of Liège, Liège, Belgium
Email: fbrouers@ulg.ac.be

## Abstract

**In this paper, we recall for physicists how it is possible using the principle of maximization of the Boltzmann-Shannon entropy to derive the Burr-Singh-Maddala (BurrXII) double power law probability distribution function and its approximations (Pareto, loglogistic.) and extension (GB2…) first used in econometrics. This is possible using a deformation of the power function, as this has been done in complex systems for the exponential function. We give to that distribution a deep stochastic interpretation using the theory of Weron *et al*. Applied to thermodynamics, the entropy *nonextensivity* can be accounted for by assuming that the asymptotic exponents are scale dependent. Therefore functions which describe phenomena presenting power-law asymptotic behaviour can be obtained without introducing exotic forms of the entropy.**

## 1. Introduction

In this paper, we want to show how the BurrrXII-Singh-Maddala (BSM) [1] [2] distribution function, known also as the $q$-Weibull distribution can be naturally derived from the maximum entropy principle using the Boltzmann-Shannon entropy with well-defined constraints including a generalization of the definition of the moment [3]-[5] similar to the deformation of the exponential. This is what has been done in Section 5. In Sec-tion 2 and 3, we recall the properties of the BSM distribution and its applications in physics and chemistry. Section 4 is devoted to summarize the Weron stochastic interpretation of the BSM probability distribution which has allowed us in Section 7 and 8 to give a novel interpretation of the concept of *nonextensivity*.

This paper has been written for several reasons.

The BSM distribution has been used in a variety apparently independent fields: econometrics [6] [7], actuary sciences [8] [9], hydrology [5], forestry [10], sorption theories [11] [12], fractal kinetics [13]-[20], pharmacokinetics [21], relaxation and reaction phenomena [22] [23]. Two of its approximations, the Weibull and the Hill equations are widely used in materials sciences [24]-[26], medical sciences in particular cancer remission and pharmacokinetic [27]-[29], physico-chemistry in particular complex and enzymatic reactions [30]-[32], meteorology [33] adsorption [34] [35], economy [36] etc. We have quoted here some of the most recent papers using the BSM, the Weibull and Hill distributions. The BSM distribution and some of the distribution derived from it give rise asymptotically to power-laws at high or (and) small values of the variable. For some values of the parameters, they belong to the family of Lévy heavy tail functions [37].

There is a widespread belief in physics that the phenomena characterized by power laws behaviour, ubiquitous in nature, should require extensions of the Boltzmann-Shannon (BS) entropy and that the BS entropy should be restricted to the family of Gaussian and exponential laws *i.e.* the distributions which obey the classical central limit theorem. Among the more than 20 proposed generalized entropies, the most famous are the Reyni [38] and the Tsallis [39] [40] entropies. To avoid divergences of the moments, Reyni introduces in the expression of the entropy powers of the probability function. Tsallis has used the same procedure introducing so-called "escort probabilities". He proposes a form of entropy which has the property to be *nonextensive* ab-initio and derives the so-called Tsallis distribution to extend the exponential distributions in thermodynamics.

We show in details, as this has been noticed by previous authors before us, that this is not necessarily true. It is possible quite naturally to obtain power law distributions by generalizing the definition of moments or using constraints determined by the data observation and keeping the BS entropy as a starting point for the maximization procedure [4] [5] [41]-[47]. M. Visser [48] claims that power laws can be obtained by applying maximum entropy ideas directly to the Shannon entropy subject only to one constraint that the logarithm of the observable quantity is specified.

The BSM cumulative distribution functions are characterized by three parameters, two form factors $a$ and $c$ and one scale parameter $b$. The power law exponents corresponding to, respectively the asymptotic behaviour of small and large values of its argument are $a$ and $\mu = a/c$. The Tsallis distribution, derived from the maximization of Tsallis entropy is the survival part of the cumulative BSM distribution when $a$ is equal to one. This density function has been used in the literature to fit many experimental data. These data can be fitted with the same degree of precision using both density distributions. This has created confusion on the value of the so-called entropic index $q$ which is different in both cases.

It must be stressed that the BSM density function includes an exponent $a$ which is absent from the Tsallis density function. The experimental observations, which dates back to the beginning of the last century [49]-[51] show that in natural phenomena, the exponent $a$ is rarely equal to one and that the relation of $\mu$ with the two parameters $a$ and $c$ is an important feature which is ignored in Tsallis formalism. The far future is not independent on the early beginning. We know it from cosmology and from all organism evolutions.

However, we will show in Section 8 that *nonextensivity* of the entropy may arise from the scale dependence of the characteristic exponents.

As a consequence of the previous points we will show that phenomena described by function with one or two tails asymptotic power laws have not necessarily to be obtained by the maximization of an extension of the BS entropy. This is has been known for a long time in fields outside of physics such as hydrology and econometrics.

## 2. The Burr-Singh-Madalla Distribution

The BSM cumulative density function is written as:

$$F_{BSM}(x;a,b,c) = 1 - \left(1 + c\left(\frac{x}{b}\right)^a\right)^{-\frac{1}{c}} \tag{1}$$

where $a$ and $c$ are form factors and $b$ is a scaling factor.

Its density function is easily obtained by differentiation

$$f_{BSM}(x;a,b,c) = \frac{a}{b}\left(\frac{x}{b}\right)^{a-1}\left(1+c\left(\frac{x}{b}\right)^a\right)^{-\frac{1}{c}-1} \quad (2)$$

It is solution of a differential equation

$$\frac{dF(x)}{dx} = f(x) = g(x)F(x)(1-F(x)) \quad (3)$$

The function $g(x) = \frac{\tilde{g}(x)}{x}$ where

$$\tilde{g}(x) = \frac{a\left(\frac{x}{b}\right)^a}{\left(1+c\left(\frac{x}{b}\right)^a\right)\left[1-\left(1+c\left(\frac{x}{b}\right)^a\right)^{-\frac{1}{c}}\right]} \quad (4)$$

The function $\tilde{g}(x)$ varies slowly from $a$ to $\mu = a/c$. One has: $\tilde{g}(x) \to a$ when $x \to 0$ and $\tilde{g}(x) \to \frac{a}{c} = \mu$, when $x \to \infty$.

The differential equation describes a birth and death process modulated by a quasi hyperbolic function which can be modified to accommodate more complex situations.

The cumulative distribution function $F_{BSM}(x)$ exhibits asymptotically two power laws: one for $x \to 0$, $F(x) \to x^a$ and one for $x \to \infty$, $F(x) \to x^{-\mu}$, $\mu = a/c$. It has a limited number of finite moment depending on the value of $\mu$.

The survival part of the cumulative function is given by

$$\left(1+c\left(\frac{x}{b}\right)^a\right)^{-1/c} \quad (5)$$

which has the same form as the Tsallis density function if $a = 1$ and $c = q-1$ where $q$ is the Tsallis entropy index. By contrast the power of the BSM density function is $-\frac{1}{c}-1 = \frac{q}{q-1}$. and the BSM density function is the so called "escort probability" in Tsallis formalism. The Lévy power law exponent $\mu$ are accordingly different

$$\mu_{BDM} = \frac{\alpha}{c} = \frac{\alpha}{q-1}\mu_T = \frac{2-q_T}{q_T-1} \quad (6)$$

and this has to be considered in interpreting experimental results.

As the mathematician Vladimir Arnold once said "Differential equations are the source of the development of modern sciences", one can consider Equation (3) as the natural starting point to study complex systems in the same way the exponential is solution of a differential equation and is the natural starting point for the description of simple systems. For instance, it has been used in epidemiology [52], the influence of the sanitary authorities to its propagation being accounted for by modifying the function g(x) accordingly. It must be reminded that the approximation g(x) = 1 is the famous Verhulst [53] equation whose discrete version is one of the paradigms in the theory of chaos [54].

## 3. Application of the BSM Distribution Function in Physics

The BSM function has been used to establish a three parameters fractal kinetic equation which has been em-ployed with some success to characterize macroscopically the sorption (ad-, chemi-, bio-) in gaseous and aqueous phase [11] [12] as well as in the theory of relaxation [22] [23] to justify the two asymptotic behaviours of the Havriliak-Negami formula [55] in the frequency range. It has been used also to show that some of the empirical isotherms (Langmuir [56], Sips-Hill [57] and Brouers-Sotolongo [34] isotherms) are well defined approximations of the BSM function and therefore enjoy the properties of statistical distributions.

The work of K. Weron and its collaborators has given a deep physical understanding of this distribution. Indeed they have derived the BSM survival function in the theory of relaxation by means of stochastic arguments linking the observed macroscopic properties to the mesooscopic and microscopic energetic and geometric organisation (fractal scaling, clustering, self-organiszation) of complex heterogeneous systems. We will deal with this interpretation in the next paragraph.

## 4. Stochastic Interpretation of the BSM Distribution

The analytical form of the BSM distribution function can be justified and the parameters $(a,b,c)$ can be given physical and statistical interpretations following the stochastic analysis given by K. Weron and collaborators [22] [58] [59] in a series of papers on relaxation and reaction in complex systems The arguments are the following: to relate macroscopic data to a macroscopic model representing the system as a whole a number of averaging formal operation at the micro- and meso-scopic level have to be done. Two cases can occur 1) the system is not strongly disordered and usual averages obeying the central limit theorem can be performed. As a consequence the probability functions belong to the basin of attraction of Gaussian functions or 2) the disorder, due to geometric and energy frustrations is giving rise to self-similar and clustering structures. As a consequence the summations of local physical quantities such as relaxation or chemical rates are dominated by their extreme values a situation well known in processes like earthquakes, water river level, atmospheric catastrophe and insurance claims to mention the most common in the literature. The corresponding distributions obey generalized limit theorems and belong to the basin of attraction of stable Lévy distributions popularized by Mandelbroot in his work on fractal structures in economics. As far as we are concerned, in that case expected values cannot be defined and specific formal methods have to be devised to tame what is called "wild disorder". One essential characteristic of these distribution functions is that they exhibit scaling properties which reveals a universal behaviour independent on the microscopic details of the system. According to Weron et al., in the time domain, the relaxation or survival function can be written as:

$$\phi(t) = \Pr(\tilde{\vartheta} \geq t) = \left\langle \exp(-t\tilde{\beta}) \right\rangle \quad (7)$$

The quantities $\tilde{\vartheta}$ and $\tilde{\beta}$ are the waiting time and the relaxation or chemical rate of a virtual macroscopic state representing the system as a whole. One has to average over random macroscopic objects which are themselves average on the mesoscopic (self-similar geometrical and dynamical clusters) and microscopic (individual reacting pairs molecules or atoms). The quantity $\tilde{\beta}$ is defined as the sum of individual relaxation rates according to equation

$$\beta = \frac{\sum_{i=1}^{N} \beta_i}{A_N} \quad (8)$$

where $A_N$ is a $N$-dependent normalizing constant.

Two situations have to be considered: the expecting value of $\tilde{\beta}$ is finite and we are in the situation of a well-behaved disorder system or it does not exist and the sum $\tilde{\beta}$ is also a random variable obeying the same probability distribution as the individual $\beta_i$ under appropriate normalization (Lévy stable distribution). If $\tilde{\beta}$ does not have a mean, the function $\phi(t)$ might have one. In that case

$$\phi(t) = \left\langle \exp(-t\tilde{\beta}) \right\rangle = \int_0^\infty \exp(-A\lambda t) g_a(\lambda) d\lambda \quad (9)$$

where $g_a(\lambda)$ is the one-sided Lévy stable density probability distribution and $A$ is a normalization constant

$$\phi(t) = \left\langle \exp(-(At))^a \right\rangle \quad (10)$$

It is the Weibull survival function quite frequent in physical, chemical and biological phenomena. The pa-rameter *a* arises from the stable scaling properties at the micro and mesoscopic levels. When $a = 1$, one has a simple exponential function and the rate is constant

The BSM distribution function can be obtained by considering a more complex situation where due to complex frustrations, the number of reacting element is not fixed and is itself a random variable. In that case Equation (8) should be replaced by

$$\tilde{\beta}^* = \sum_{i=1}^{v_N} \frac{\beta_i}{A_N} \tag{11}$$

As argued by Weron *et al*. the fluctuations of $v_N$ can be view as a birth and death process. In that case the most natural probability distribution is the negative binomial probability distribution which in the limit $N \to \infty$ tends to the gamma distribution:

$$\Gamma\left(\lambda, c, \frac{1}{c}\right) = \frac{1}{c\Gamma\left(\frac{1}{c}\right)} \left(\frac{\lambda}{c}\right)^{\frac{1}{c}-1} \exp\left(-\frac{\lambda}{c}\right) \tag{12}$$

Then the survival probability function of the entire system is

$$\phi(t) = \left\langle \exp\left(-t\tilde{\beta}^*\right)\right\rangle = \int_0^\infty \exp\left(-\lambda(At)^a\right) \Gamma\left(\lambda, c, \frac{1}{c}\right) d\lambda \tag{13}$$

The solution of which has been obtained by Rodriguez [60] more than forty years ago

$$\phi(t) = \left(1 + c(At)^a\right)^{-1/c} = 1 - F_B\left(a, A^{-1}, c\right) \tag{14}$$

This is the BSM survival distribution function in the time domain. The corresponding density function is

$$f(t) = aA(At)^{a-1}\left(1 + c(At)^a\right)^{-\frac{1}{c}-1} \tag{15}$$

In this derivation of the BSM function, the parameter *c* appears to be a measure of aggregation with one or several characteristic lengths.

## 5. The BSM Density Function Derived from the Maximization of the BS Entropy

Starting from the Boltzmann-Shannon entropy

$$S = -\int_0^\infty f(x;a,b,c) \ln\left(f(x;a,b,c)\right) dx \tag{16}$$

We determine $f(x;a,b,c)$ by imposing three constraints

$$\int_0^\infty f(x;a,b,c) dx = C_1 \tag{17}$$

$$\int_0^\infty \ln\left(\frac{x}{b}\right) f(x;a,b,c) = \int_0^\infty \varphi_2(x) f(x;a,b,c) dx = C_2 \tag{18}$$

And generalizing the definition of the power of a variable in the same spirit as the definition of the deformed exponential

$$x_c^a = \frac{1}{c}\ln\left(1 + cx^a\right) \quad \text{with } x_0^a = x^a \tag{19}$$

The third constraint can then be expressed as:

$$\int_0^\infty \ln\left(1 + \left(\frac{x}{b}\right)^a\right) f(x;a,b,c) dx = \int_0^\infty \varphi_3(x) f(x;a,b,c) dx = C_3 \tag{20}$$

The constraint conditions can be understood as known prior information which can be used to achieve a least biased distribution.

Using the method of Lagrange optimization method and defining

$$\mathcal{L} = S(x) - \sum_{j}^{3} \left( \lambda_j \int_0^\infty \varphi_j(x) f(x;a,b,c) dx - C_j \right) \quad (21)$$

The maximization of $S(x)$ is obtained by solving the equation: $\frac{d\mathcal{L}}{df(x)} = 0$ which taking account of the normalization condition $C_1 = 1$ yields the functional form:

$$f(x;a,b,c) = \frac{1}{Z(\lambda_1,\lambda_2)} \exp\left[-\lambda_2 \varphi_2(x) - \lambda_3 \varphi_3(x)\right] \quad (22)$$

With the normalization constant determined by the partition function

$$Z(\lambda_1,\lambda_2) = \int_0^\infty \exp\left[-\lambda_2 \varphi_2(x) - \lambda_3 \varphi_3(x)\right] dx \quad (23)$$

The values of the $\lambda_j$ parameters are determined by the set of equations

$$C_j = \frac{\partial \log Z(\lambda_2,\lambda_3)}{\partial \lambda_j} \quad \text{and} \quad \frac{\partial S(x)}{\partial \lambda_j} = \int_0^\infty f(x)\phi_j(x) dx = C_j \quad (24)$$

one gets finally, given the constraints

$$C_2 = -\frac{1}{a}\left(\gamma + \ln(cb^a) - \psi\left(\frac{1}{c}\right)\right), \quad C_3 = c \quad (25)$$

$\gamma$ is the Euler constant and $\psi(x)$ the Bigamma function.

We have therefore

$$f(x) = \frac{1}{Z}\left(\frac{x}{b}\right)^{-\lambda_2} \left(1 + \left(\frac{x}{b}\right)^a\right)^{-\lambda_3} \quad (26)$$

with

$$\lambda_2 = 1-a, \quad \lambda_3 = 1 - \frac{1}{c}, \quad Z = \frac{b}{a} \quad (27)$$

This finally gives the BSM density function (Equation (2)).

The expression for the $k$-th moment is given by [61]

$$\langle x^k \rangle = \frac{b^k}{c^{k/a}} \frac{\Gamma\left(1+\frac{k}{a}\right)\Gamma\left(\frac{1}{a} - \frac{k}{a}\right)}{\Gamma\left(\frac{1}{c}\right)}, \quad k < \frac{a}{c} \quad (28)$$

The Tsallis density function ($a = 1$) is the survival function of $F(x;1,b,c)$

$$f_T(x;1,b,c) = \left(1 + c\left(\frac{x}{b}\right)\right)^{-\frac{1}{c}} \quad (29)$$

with the constraints

$$C_2 = -\gamma - \ln(cb) + \Phi\left(\frac{1}{c}\right), \quad C_3 = c \quad (30)$$

These results have already been obtained with other notations in the field of econometrics and meteorology [3]-[5] [45]. The same procedure has been used also for 4-parameters generalization of the BSM (GB2 foe example) in the econometrics literature. One recovers our results when the extra parameter is put to one or zero according to the type of extension.

Another tail distribution, the Cauchy distribution $\frac{1}{\pi(1+x^2)}$, one of the Levy distributions has been derived with the same method with the constraint

$$\int_{-\infty}^{\infty} \ln(1+x^2) f(x) dx = 2\ln 2 \qquad (31)$$

The Weibull and the Pareto and Cauchy tail distribution entropies are well known and have been derived in the classical books on entropy maximization method [62].

For $c = 0$, one recovers the well-known results for the Weibull distribution

$$\int_0^{\infty} \ln x f_W(x;a,b) dx = -\frac{\gamma}{a} \qquad (32)$$

$$\int_0^{\infty} \left(\frac{x}{b}\right)^a f_W(x;a,b) dx = 1 \qquad (33)$$

with

$$f_W(x;a,b) = f_B(x;a,b,0) = \frac{ax^{a-1}}{b^a} \exp\left(-\left(\frac{x}{b}\right)^a\right) \qquad (34)$$

For $c = 1$, the log-logistic-Hill-Fisk constraints

$$\int_0^{\infty} \ln x f_H(x;a,b) dx = b \int_0^{\infty} \left(\frac{x}{b}\right)^a f_H(x;a,b) dx = 1 \qquad (35)$$

with

$$f_H(x;a,b) = f_B(x;a,b,1) = \frac{ax^{a-1}}{b^a}\left(1+\left(\frac{x}{b}\right)^a\right)^{-2} \qquad (36)$$

Finally it must be noted that if $x \gg b$ (Zipf law), the two constraints (18) and (20) reduced to only one on the logarithm and we reach the same conclusion as Visser [48].

## 6. Expressions of the Entropy

Making use of the results of the previous section, one can now write the expressions of the entropy corresponding to the various approximations.

From the general definition

$$S(x) = -\int_0^{\infty} f_B(x) \ln(f_B(x)) dx \qquad (37)$$

We get for the BSM distribution

$$S_B(a,b,c) = \ln\left(\frac{b}{a}\right) + (a-1)\frac{\gamma + \ln(c) - \psi\left(\frac{1}{c}\right)}{a} + c + 1 \qquad (38)$$

for the Weibull distribution

$$S_B(a,b,0) = \ln\left(\frac{b}{a}\right) + (a-1)\frac{\gamma}{a} + 1 \qquad (39)$$

For the log-logistic-Hill-Fisk distribution

$$S_B(a,b,1) = \ln\left(\frac{b}{a}\right) + 2 \qquad (40)$$

and

$$S_B(1,b,c) = \ln b + c + 1 \qquad (41)$$

$$S_B(1,b,1) = \ln b + 2 \qquad (42)$$

$$S_B(1,b,0) = \ln b + 1 \tag{43}$$

The general form of the entropy is therefore

$$S(x) = \log Z(x) + K(a,c) \tag{44}$$

$K(a, c)$ is a constant that we can consider as the origin of the entropy for a couple of values $a$ and $c$. The previous results are therefore compatible with the *extensivity* of the BS entropy. If we have two subsystems 1 and 2 added to form a larger system with have:

$$\ln(Z) = \ln(Z_1 Z_2) = \ln Z_1 + \ln Z_{21} \tag{45}$$

This is only true if $(a,c) = (a_1,c_1) = (a_2,c_2)$. Otherwise the system is *nonextensive*. We will come back to this situation in the two next sections.

## 7. Implications in Thermodynamics

The aim of this paper is to discuss the probabilistic and stochastic foundation of the use of the BSM distribution to describe physical and chemical complex systems characterized by power-law, Levy and extreme value distributions. We will nevertheless in the last section touch the problem of the application of this formalism to thermodynamics.

In the case ($a = 1$, $c = 0$, $q = 1$), we have $S = \ln b + 1$, in the canonical version of thermodynamics the $x$ variable can be replaced by the individual particle energy $\varepsilon$ and we can obtain in the continuous limit the set of relations

$$f(x) = \frac{\exp\left(-\dfrac{\varepsilon}{b}\right)}{Z}, \quad Z = \int_0^\infty \exp\left(-\frac{\varepsilon x}{b}\right)dx = b, \quad \langle \epsilon \rangle = b \tag{46}$$

The corresponding canonical entropy is

$$S = \ln Z + \frac{\langle \epsilon \rangle}{kT} \tag{47}$$

In the case $a = 1$, which is the case used in extensions of the classical thermodynamics, one has

$$S(a=1,b,c) = \ln b + q, \quad q = c+1 \tag{48}$$

$$f_B(\varepsilon;1,b,c) = \frac{1}{b}\left[1+c\left(\frac{\varepsilon}{b}\right)\right]^{-\frac{c+1}{c}} = \frac{1}{b}\left(\exp_q\left(-\frac{\varepsilon}{b}\right)\right)^q \tag{49}$$

$$b = \int_0^\infty \left(\exp_q\left(-\frac{\varepsilon}{b}\right)\right)^q d\varepsilon = Z_q \tag{50}$$

$$\langle \epsilon \rangle = \frac{b}{1-c} = \frac{b}{2-q} \tag{51}$$

$$S(a=1,b,c) = \ln Z_q + \frac{q(2-q)}{b}\langle \epsilon \rangle, \quad q=c+1 \tag{52}$$

$$= \ln Z_q + \frac{(1-c^2)}{b}\langle \epsilon \rangle \tag{53}$$

Therefore if $\langle \epsilon \rangle$ is constant, entropy decreases when $c$ increases.

The usual canonical thermodynamics expression for the entropy (Equation (47)) can be recovered if one introduces a $c$-temperature assuming that $T_c$ is proportional to $\langle \epsilon_c \rangle$:

$$1+c = \frac{\langle \epsilon_c \rangle}{T_c} = \frac{Z_c}{(1-c)T_c} \tag{54}$$

which gives a *c*-depending temperature depending on the evolution of $Z_c$ (and therefore on $S_c$) with *c*.

$$T_c = \frac{Z_c}{(1-c^2)} = \frac{Z_q}{q(2-q)} = b \tag{55}$$

## 8. *Nonextensivity*

*Nonextensivity* observed in systems with long range interactions can be accounted for if the parameter *c* (and *q*) are scale dependent. One can use an argument used by C. Beck [63] to obtain a quasi nonexpensive entropy from a superstatistic version of the Tsallis entropy. I quote "in other words *q*(*r*) is a strictly monotonously decreasing function of the scale, just as observed in experiments". This means that, in the BSM formalism, if the parameter *c* decreases with an increase of the volume; the additive property of the entropy is no longer respected since

$$S(V,c) \neq S(V_1,c_1) + S(V_2,c_2), V = V_1 + V_2, \quad c < c_1, c < c_2 \tag{56}$$

Moreover, the system is no longer in an equilibrium state since $\langle \epsilon_c \rangle$ and the temperature $T_c$ are also *c*-dependent.

This interpretation of *nonextensivity* is in agreement with the signification of the parameter *c* which is related to the cluster organization (possibly multifractal) of the heterogeneous system as discussed in Section 4.

## 9. Conclusion

In conclusion, we think that physicists have to learn a lot from the progress made the last decades by mathematicians and ex-physicists in the field of statistics in econometrics. We think that many phenomena exhibiting power-law behavior are not necessarily the consequence of a *nonextensivity* of the entropy as this has been assumed by a number of authors, myself [64]-[66] included, evoking exotic form of the entropy. The same results can be obtained using the BSM density function and taking account of the difference in the definition of the exponents (Equation (6)). In our presentation *nonextensivity*, when it occurs, it is the consequence of the scale dependence of the characteristic exponents *a* and *c* of the distribution and is a property of systems with long range interactions and complex systems, where thermodynamic equilibrium is not achieved.